\documentclass[11pt]{article}

\usepackage[final]{acl}

\usepackage{times}
\usepackage{latexsym}
\usepackage{amsmath}
\usepackage{pifont}
\usepackage{amssymb}
\usepackage{hyperref} % 导入超链接包
\usepackage{xcolor}   % 可选，用于颜色设置

\usepackage{amssymb}
\usepackage{booktabs} % 用于更好的表格线
\usepackage{multirow} % 用于多行合并
\usepackage{array}    % 用于列格式调整
\usepackage{caption}  % 用于标题格式
\usepackage{makecell}
\usepackage{pdflscape, geometry}
\usepackage{graphicx}
\usepackage{subcaption} % 提供 subfigure 环境
\usepackage[T1]{fontenc}
\usepackage{tikz}
\newcommand*\circled[1]{\tikz[baseline=(char.base)]{
            \node[shape=circle,draw,inner sep=1pt] (char) {#1};}}

\usepackage[utf8]{inputenc}

\usepackage{microtype}

\usepackage{inconsolata}

\usepackage{graphicx}

\title{DisCo-Speech: Controllable Zero-Shot Speech Generation with A Disentangled Speech Codec}

\author{
  Tao Li\thanks{ Equal contribution. \textsuperscript{\dag} Corresponding authors.}, 
  Wenshuo Ge\textsuperscript{*},
  Zhichao Wang,
  Zihao Cui,
  Yong Ma, \\
  \textbf{Yingying Gao},
  \textbf{Chao Deng},
  \textbf{Shilei Zhang}\textsuperscript{\dag},
  \textbf{Junlan Feng}\textsuperscript{\dag}
  \\
  \\
   China Mobile Nineverse Artificial Intelligence Technology (Beijing) Co., Ltd. \\
   Nineverse  Institute of Artificial Intelligence.\\
   The State Key Laboratory of Multimedia Information Processing, Peking University, Beijing, China. 
}

\begin{document}
\maketitle
\begin{abstract}
Codec-based language models (LMs) have revolutionized text-to-speech (TTS). However, standard codecs entangle timbre and prosody, which hinders independent control in continuation-based LMs. To tackle this challenge, we propose DisCo-Speech, a zero-shot \textbf{co}ntrollable TTS framework featuring a \textbf{dis}entangled speech codec (DisCodec) and an LM-based generator.
The core component DisCodec employs a two-stage design: 1) tri-factor disentanglement to separate speech into content, prosody, and timbre subspaces via parallel encoders and hybrid losses; 
and 2) fusion and reconstruction that merges content and prosody into unified content-prosody tokens suitable for LM prediction, while jointly optimizing reconstruction to address the disentanglement-reconstruction trade-off. 
This allows the LM to perform prosodic continuation from a style prompt while the decoder injects target timbre, enabling flexible zero-shot control. Experiments demonstrate that DisCo-Speech achieves competitive voice cloning and superior zero-shot prosody control. 
By resolving the core entanglement at the codec level, DisCo-Speech provides a robust foundation for controllable speech synthesis. 
Audio samples are available at: \url{https://disco-speech.github.io/DisCo-demo/}. Code and weights will be released at: \url{https://github.com/disco-speech/DisCo-Speech} upon acceptance.

\end{abstract}

\section{Introduction}
\label{introduction}

Text-to-speech (TTS) synthesis, the technology that converts written text into spoken audio, has long been the core of human-computer interaction~\cite{tacotron,fastspeech}. 
Recently, TTS has witnessed a paradigm shift with the rise of codec-based language models, in which codecs discretize speech into tokens via quantization techniques~\cite{vqvae,fsq}, and language models (LMs) bridge the correlation between text and speech. 
Driven by large-scale datasets and advanced generative models, zero-shot voice cloning, where a new speaker's voice can be cloned with a single speech clip, is no longer a barrier in modern TTS systems~\cite{seedtts,cosyvoice2}.

With the proliferation of TTS applications, a new demand has emerged for fine-grained and independent control over speech attributes, such as speaker timbre and prosody, enabling a target speaker to speak in any desired prosody—a task referred to as \textit{zero-shot controllable} speech generation~\cite{ustyle,vevo,indextts2}.
However, relying on existing acoustic~\cite{soundstream,xcodec} or hybrid codecs~\cite{speechtokenizer,mimi} presents a challenge: the inherent tight coupling of timbre and prosody within these representations hinders current LM-based TTS systems~\cite{recent,minimax} from meeting this requirement.
Although the continuation-based generation paradigm of LMs excels at high-similarity cloning by replicating both timbre and prosody from the speech prompt, it inevitably sacrifices the capability for independent control.

To break this entanglement, one intuitive approach involves building comprehensive, multi-style speaker datasets with fine-grained prosodic annotations, allowing TTS systems to explicitly learn prosody and timbre from separate prompts~\cite{unisyn}.
However, the high resource consumption makes this approach difficult to scale, especially in zero-shot scenarios.
Alternatively, many efforts~\cite{naturalspeech3,msrcodec,freecodec} focus on the design of codec, aiming to provide disentangled speech attribute tokens~(e.g., content, prosody, and timbre), yet decoupling remains a critical bottleneck. 
The trade-off between disentanglement and reconstruction~\cite{diclet} often leads to information loss or leakage~\cite{copycat}, compromising both synthesis quality and the effectiveness of independent control.

Achieving effective disentanglement in codecs to promote zero-shot controllable TTS is non-trivial, primarily due to several core challenges in speech representation modeling:

$\bullet$~\textbf{Speech disentanglement dilemma}: among speech attributes, timbre is globally static, and content and prosody are temporal dynamics~\cite{leicross,yuepeng,licon}. 
Content is a form of linguistic information~\cite{ustyle}. 
Prosody can encompass high-level content-independent style and further impact the tone and intensity attached to the content, while speaker timbre further adjusts the prosody, forming diverse human-observed expression~\cite{diclet}. 
This hierarchy is evidenced in layer-wise analyses of SSL or ASR models~\cite{vevo,layerwise,streaming,distilhubert}, where timbre information fades before prosody as layers deepen. Strict decoupling risks disrupting these intrinsic dependencies causing information loss, while weak constraints permit information leakage~\cite{ltinter,unisyn}.

$\bullet$~\textbf{Disentanglement-reconstruction trade-off}: 
the trade-off between disentanglement and reconstruction has been reported in previous studies~\cite{unet}. 
Excessive disentangling can strip away acoustic details essential for high-fidelity synthesis. 
Conversely, prioritizing reconstruction quality often leaves entangled information in the representations, limiting the precision of downstream control~\cite{ustyle}.

$\bullet$~\textbf{Downstream-friendly representation}: a practical disentangled representation is not only sufficiently pure but also convenient for the usage of downstream components (e.g., LMs)~\cite{recent,vevo}. How to construct a representation that is easily utilizable by a zero-shot control framework is key to unlocking the codec's potential~\cite{vevo,indextts2}.

Addressing these challenges, we propose \textbf{DisCo-Speech}, a novel framework for zero-shot \underline{\textbf{co}}ntrollable speech generation, comprising a \underline{\textbf{dis}}entangled speech codec (\textbf{DisCodec}) and a single Transformer LM.
At the core of our framework lies DisCodec, which facilitates independent zero-shot control through a two-stage learning paradigm:
1) Tri-factor disentanglement: Inspired by the characteristic of speech attributes as mentioned above, DisCodec factorizes speech into content, prosody, and timbre via three parallel encoders,
employing hybrid constraints to ensure robust disentanglement;
2) Fusion and reconstruction: A token to wave decoder fuses the disentangled content and prosody into unified, timbre-agnostic tokens suitable for LM prediction, while jointly optimizing reconstruction to mitigate the disentanglement-reconstruction trade-off.
By resolving entanglement at the codec level, the LM and DisCodec decoder seamlessly collaborate: the LM performs contextual prosodic continuation based on the text and prosody prompt, while the decoder reconstructs the waveform conditioned on the target timbre. 
This design establishes DisCo-Speech as a concise and effective paradigm for independent zero-shot control.

\begin{figure*}[htb]
  \includegraphics[width=0.9\linewidth]{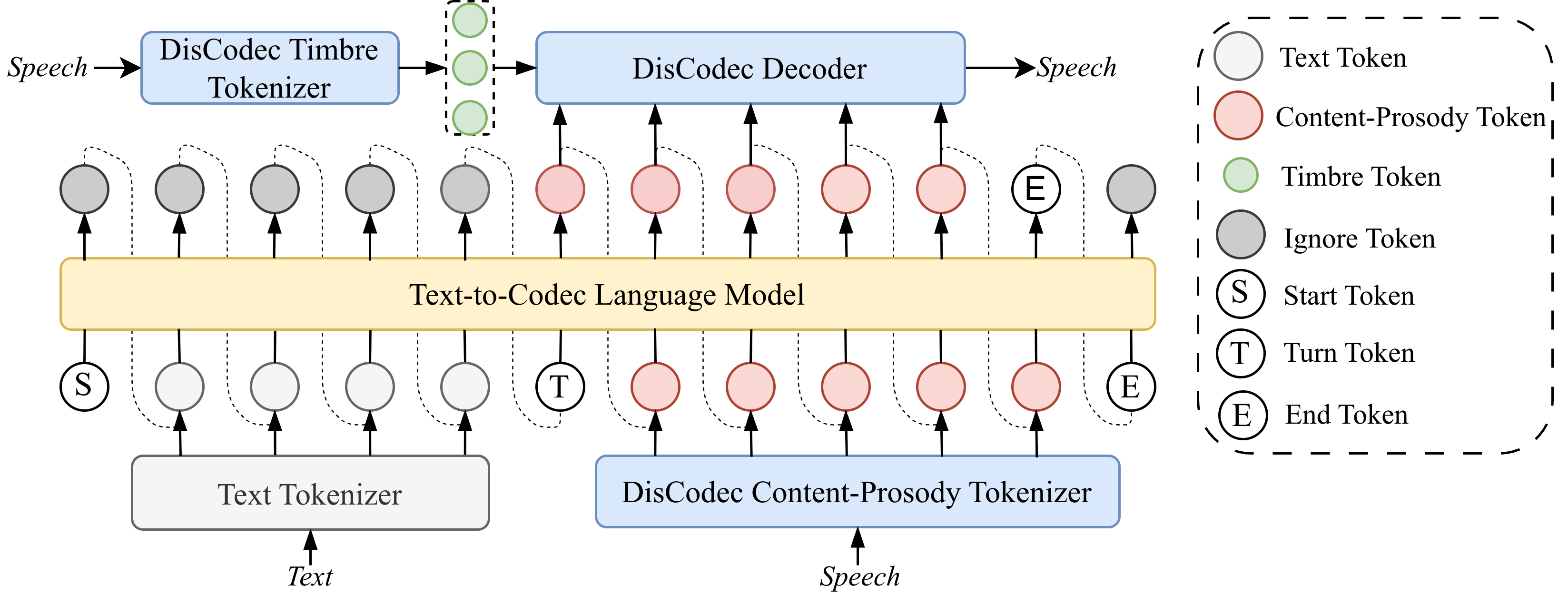}
  \centering
  \caption {The overview of DisCo-Speech.}
  \label{fig:disco}
  \vspace{-14pt}
\end{figure*}

\section{Related Works}

\subsection{Speech Tokenization}

The discrete codec paradigm, built upon an encoder-quantizer-decoder architecture~\cite{vqvae}, has become the foundation for modern TTS, enabling speech representation compatible with language models. Acoustic codecs~\cite{soundstream,encodec,bigcodec,xcodec} established this paradigm, focusing primarily on acoustic representations and reconstruction quality through residual vector quantization~\cite{rvq} (RVQ) or finite scalar quantization~\cite{fsq} (FSQ). 
To bridge text-speech modality gap, ASR- or SSL-based semantic codec~\cite{cosyvoice1,cosyvoice2,seedtts} have been introduced into TTS frameworks to improve generation stability. 
Benefiting from semantic and acoustic modeling, semantic-aware acoustic codec~\cite{mimi,speechtokenizer} 
have emerged as mainstream solutions, progressively representing speech from semantic to acoustic levels via semantic distillation within a multi-layer structure.

A recent frontier lies in disentangled codecs, which factorize speech into distinct attributes (e.g., content, prosody, and timbre) for fine-grained control.
Several approaches have been explored: FACodec~\cite{naturalspeech3} employs gradient reversal layers~\cite{grl} (GRL) and multi-aspect supervision; MSR-Codec~\cite{msrcodec} and FreeCodec~\cite{freecodec} leverage pre-trained models to decouple attributes. 
Despite these advancements, insufficient decoupling often hampers controllable performance~\cite{controlspeech}, and many methods rely on specialized acoustic models~\cite{naturalspeech3}, complicating the pipeline. 
We propose DisCodec that provides well-disentangled representations for seamless integration with standard LMs, yielding a more concise and effective pipeline.

\subsection{Zero-shot Controllable TTS}
Zero-shot controllable TTS aims to synthesize speech with desired speaker timbre and speaking prosody, using acoustic prompts or textual instructions. 
Early zero-shot TTS systems~\cite{valle,speartts,xtts,e2tts} focused primarily on voice cloning, where a single acoustic prompt jointly defines both speaker timbre and speaking style. 
For instance, VALL-E~\cite{valle} introduced an AR+NAR architecture that leverages in-context learning to replicate timbre from an acoustic prompt. 
Subsequent works have improved modeling capability through progressive semantic-to-acoustic modeling~\cite{cosyvoice2,seedtts} or integrated diffusion-based hybrid architectures~\cite{cosyvoice1,minimax,ditar}.

As cloning performance improved, the need for prosody control arose. 
An intuitive approach involves using annotated multi-style, multi-speaker data. Textual instructions~\cite{cosyvoice1,promptstyle,controlspeech,instructtts} or acoustic templates~\cite{stepaudioedix} are often employed to guide style control. 
Due to the high annotation cost, some studies attempt to factorize speech into content, timbre, and prosody to achieve independent control.
Recently proposed IndexTTS2~\cite{indextts2} incorporates a GRL-based disentanglement module trained jointly with an LM to separately control timbre and style, while Vevo~\cite{vevo} and NaturalSpeech3~\cite{naturalspeech3} leverage disentangled features from pre-trained SSL models or codecs. 
Despite these advances, existing methods still face challenges like insufficient decoupling and reliance on specialized architectures, limiting their flexibility ~\cite{recent,controlspeech}, our DisCo-Speech offers a more versatile and concise solution that leverages a disentangled codec (DisCodec) to enable a standard LM to independently control timbre and prosody.

\section{DisCo-Speech}

As illustrated in Fig.~\ref{fig:disco}, DisCo-Speech comprises two core components: 1) \textbf{DisCodec}: which tokenizes speech into content-prosody and global timbre tokens and reconstructs them into a waveform;
2) \textbf{Text-to-Codec LM}: a standard LM that autoregressively generates content-prosody tokens given text and historical content-prosody tokens.
During inference, using a speech prompt with desired prosody and its corresponding text as prompts, the LM performs prosodic continuation on the target text to generate the content-prosody tokens. The generated results, together with the target speaker's timbre, are then processed by the DisCodec decoder to produce the final speech.
In the following sections, we introduce the construction of a disentangled codec suitable for a zero-shot controllable framework, and detail how a standard LM collaborates with DisCodec to achieve independent control, forming the DisCo-Speech framework.

\begin{figure*}[htb]
    \centering
  \includegraphics[width=1\linewidth]{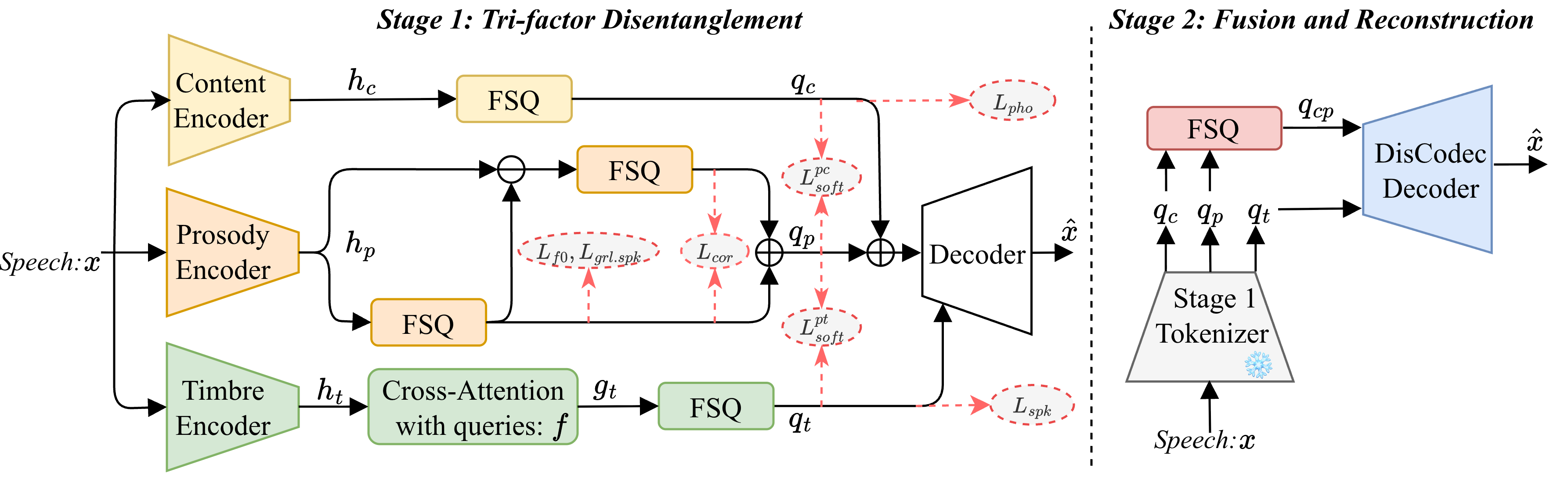}
  % \vspace{-5pt}
  \caption {The structure and two-stage training of DisCodec.}
  \label{fig:DisCodec}
  \vspace{-15pt}
\end{figure*}

\subsection{DisCodec: Disentangled Speech Codec}

As discussed in Section \ref{introduction}, the inherent inter-dependencies of speech attributes can lead to failures in either generation quality or attribute control. 
Furthermore, the intrinsic disentanglement-reconstruction conflict and the need for downstream-friendly~\cite{multitoken} feature design impose additional requirements for codec design. 
To overcome these challenges, DisCodec is designed with a two-stage training paradigm, including: 1) Tri-factor disentanglement: This stage explicitly decouples speech into content, prosody, and timbre under the guidance of hybrid decoupling constraints, ensuring the integrity of each attribute; 2) Fusion and reconstruction:
The DisCodec decoder further fuses content and prosody into unified tokens suitable for standard LM usage, while jointly optimizing reconstruction quality to directly mitigate the disentanglement-reconstruction trade-off.

\subsubsection{Stage 1: Tri-factor Disentanglement}
As shown in Fig.~\ref{fig:DisCodec}, in stage 1, three parallel encoders are employed to capture content~$c$, timbre~$t$, and prosody~$p$ from speech, and subsequently, FSQ-based quantizers perform discretization.
According to the inherent characteristics of speech attributes, varied decoupling constraints are imposed on different attribute branches for clear disentanglement. 
Additionally, with former discrete tokens, a decoder performs a reconstruction task to provide reconstruction supervision.
Given a speech $x$, the process of DisCodec in stage 1 can be described as:
\vspace{-5pt}
\begin{equation}
% \normalsize
  \label{eq:example} % 整个块共用一个编号
  \begin{aligned}
    h_c &= E_c(x), h_t = E_t(x), h_p = E_p(x),  \\
    g_t &= CrossAttention(h_t, f), \\
    q_c &= Q_c(h_c), q_p = RQ_p(h_p), q_{t} = Q_t(g_t), \\
    \hat{x} &= D(q_c, q_p, q_{t}),
  \end{aligned}
\end{equation}
where $E_c$, $E_t$, and $E_p$ are the content, timbre, and prosody encoders; $Q$ means the quantizer, while $RQ$ represent residual version; $f$ is a set of learnable queries; and $D$ is the decoder.

\textbf{Content Tokenizer.} The content encoder $E_c(\cdot)$ follows the design of the DAC encoder~\cite{dac}, which employs several convolutional blocks to downsample waveform $x$ into frame-level latent $h_c$. And then $h_c$ is quantized by FSQ $Q_c(\cdot)$ to the quantized embedding $q_c$.

To ensure $q_c$ exclusively encodes content information, a finetuned Wav2Vec-based phone recognition model~\cite{wav2vec} is employed to provide phonetic supervision, where $q_c$ are passed through a classifier to learn phone prediction under the CE-based guidance $L_{pho}$ of the recognition model. 
Instead of using SSL models~\cite{hubert}, utilizing a phone- or text-based model provides purer content supervision, lowering decoupling complexity in the codec.
%cross entropy

\textbf{Prosody Tokenizer.} To capture temporal variations of prosody, the prosody encoder $E_p(\cdot)$ employs dilated causal convolutions~\cite{wavenet} to produce frame-level sequence $h_p$. 
Unlike the content branch, a two-layer residual FSQ~($RQ_p$) is used to quantize $h_p$ to the residual-enhanced representation $q_p$, which integrates the quantized information from both FSQ layers. 
The key innovation in this design is the hierarchical assignment of prosodic attributes: the first FSQ layer is forced to encode the primary prosody attribute (i.e., pitch information), while the second FSQ in the residual path captures prosodic residuals beyond pitch for comprehensive prosody modeling.

To supervise the \textbf{prosody capturing}, the first FSQ layer is updated with a frame-level F0 regression loss $\mathcal{L}_{f0}$, and correlation loss $\mathcal{L}_{\text{cor}}$ ensures the correlation of quantized results from two FSQ layers to force the second FSQ encode prosody-related information from the residual, which can be defined as:
\begin{equation}
\label{eq2}
\small
    \mathcal{L}_{\text{cor}} = \left( \frac{1}{BL} \sum_{b=1}^{B} \sum_{l=1}^{L} \frac{q_{p1}^{(b,l)} \cdot q_{p2}^{(b,l)}}{\|q_{p1}^{(b,l)}\| \cdot \|q_{p2}^{(b,l)}\|} - \alpha \right)^2,
\end{equation}
where $B$ is the batch size, $L$ is the sequence length, $q_{p1}$ and $q_{p2}$ are the quantized output of the first and second FSQ layers, respectively, and $\alpha$ is a target similarity value (set to 0.2) that promotes moderate correlation between the two layers' representations. 
To eliminate speaker timbre, the widely used GRL layer~\cite{grl} is employed in the first FSQ layer. 
Moreover, to further ensure the \textbf{speech attribute decoupling}, inspired by the inherent relationship among speech attributes (See Section~\ref{introduction}), we introduce \textit{soft orthogonality constraint} $\mathcal{L}_{soft}$ that strikes a balance between feature decoupling and information preservation via \textit{adjustable decoupling coefficient} $\beta$. 
This soft constraint is applied to prosody-content and prosody-timbre decoupling, forming $\mathcal{L}_{soft}^{p,c}$ and $\mathcal{L}_{soft}^{p,t}$, which are described as:
\vspace{-5pt}
\begin{equation}
\label{eq34}
\small
\mathcal{L}_{soft}^{p,c} = \left( \frac{1}{BL} \sum_{b=1}^{B} \sum_{l=1}^{L} |\cos(l_p^{(b,l)}, l_c^{(b,l)})| - \beta_c \right)^2,
\end{equation}
\begin{equation}
\small
\mathcal{L}_{soft}^{p,t} = \left( \frac{1}{BL} \sum_{b=1}^{B} \sum_{l=1}^{L} |\cos(l_p^{(b,l)}, q_t^{(b)})| - \beta_t \right)^2,
\end{equation}
where $l_p$ and $l_c$ are the linear-transformed results of quantized prosody and content, respectively. 
Unlike hard orthogonality constraint~\cite{taslpemo} (i.e., $\beta\rightarrow0$ ) which pose a strict independence assumption which may cause excessive information loss, our soft version constraints strike a balance via the adjustable coefficient $\beta$. Compared with the strict decoupling ($\beta_t=0.0001$) in $\mathcal{L}_{soft}^{p,t}$, the relative higher value ($\beta_c=0.01$) in $\mathcal{L}_{soft}^{p,c}$ acknowledges the natural temporal-dynamic correlation between prosody and content, while the entanglement of timbre and prosody, reflected in coarse granularity, can achieve near-complete independence.

\textbf{Timbre Tokenizer.} Following previous studies~\cite{sparktts}, a sequence of fixed-length global tokens is used to capture the global speaker timbre. Specifically, the timbre encoder $E_t(\cdot)$ follows ECAPA-TDNN~\cite{ecapa-tdnn} to produce frame-level representations $h_t$. These are then aggregated into a fixed-length sequence $g_t$ via cross-attention with learnable queries $f$, thereby adaptively focusing on global-consistency timbre information. 
A FSQ layer $Q_t(\cdot)$ further perform quantization to produce global timbre representation $q_{t}$, implicitly creating an information bottleneck to discard non-timbre information.
To ensure effective \textbf{speaker timbre modeling}, we directly optimize the timbre tokenizer with a speaker classification loss $\mathcal{L}_{spk}$, while the soft orthogonal constraint $\mathcal{L}_{soft}^{p,t}$ mentioned above further eliminates prosodic variations from the timbre representation.

Finally, the decoder $D(\cdot)$, mirroring the architecture of the content encoder, recombines the triple stream representation back to the waveform. Multi-scale Mel-spectrogram loss and waveform reconstruction loss~\cite{dac} are employed to guide the reconstruction.
Note that the decoder is only used in stage 1.

\subsubsection{Stage 2: Fusion and Reconstruction}
In Stage 1, DisCodec achieves tri-factor disentanglement of speech attributes. 
However, the three-stream representation is not well-suited for downstream tasks such as controllable generation, as it requires predicting multiple token streams~\cite{multitoken}. 
Moreover, the inherent disentanglement-reconstruction trade-off limits the reconstruction quality of the Stage 1 decoder. 
To bridge these gaps, as illustrated in Fig.~\ref{fig:DisCodec}, we introduce a specialized decoder to optimize reconstruction quality while keeping the encoders frozen. 
To improve downstream usability, and inspired by the inherent relationship between content and prosody, we first sum the quantized embeddings of content and prosody, and then re-quantize the fused result into a unified token sequence $z_{cp}$.
The corresponding quantized embeddings $q_{cp}$ are then consumed by the decoder to reconstruct the waveform, conditioned on the global speaker timbre $q_t$, where this entire process is jointly optimized.
This design decomposes the DisCo-Speech framework into two clear steps: prosodic continuation on text, followed by timbre injection—a structure aligned with prior studies~\cite{ustyle,vevo}.

Regarding the architecture, the decoder stacks Transformer blocks~\cite{atten} with a BigVGANv2~\cite{bigvganv2} generator. 
During Stage 2, the updated DisCodec is trained with the original loss in BigVGANv2, including multi-scale reconstruction losses~\cite{hifigan}, feature matching loss~\cite{melgan}, and adversarial loss~\cite{bigvganv2}. 
In zero-shot controllable inference, the DisCodec decoder directly synthesizes the waveform from the LM-predicted content-prosody tokens $z^{sys}_{cp}$, conditioned on the target speaker’s timbre $q^{trg}_t$.

The total loss is a weighted sum of the component losses. Detailed loss hyperparameters (e.g., $\lambda_{pho}, \lambda_{cor}, \lambda_{spk}$) are given in Appendix~\ref{aploss}.

\subsection{Text-to-Codec Language Model}
Building upon the disentangled representations from DisCodec, we employ a standard LM as the generative core of DisCo-Speech. 
As shown in Fig.~\ref{fig:disco}, the LM is responsible for learning the relationship between text and prosody and generating the timbre-agnostic content-prosody token $z_{cp}$.

\textbf{Training.} During training, the input sequence is structured as: $[\circled{S}, t_{c}, \circled{T}, z_{cp}, \circled{E}]$, where $t_c$ is the byte pair encoding~(BPE) sequence of text, $z_{cp}$ is the unified content-prosody tokens from DisCodec, and \texttt{S}, \texttt{T}, \texttt{E} are special tokens indicating the start, turn, and end of the sequence. 
The model is trained with next token prediction mechanism with pretraining and supervised fine-tuning (SFT) process.

\textbf{Inference.} 
The input is constructed as: $[\circled{S}, t_{c}^{prompt}, t_{c}^{sys}, \circled{T}, z_{cp}^{prompt}]$, where $t_c^{prompt}$ and $z_{cp}^{prompt}$ are extracted from the prompt speech with the desired prosody, and $t_c^{sys}$ is the target text to be synthesized. The LM generates $z_{cp}^{sys}$ by capturing the prosodic pattern from the prompt $z_{cp}^{prompt}$. 
The final waveform is synthesized by the DisCodec decoder from $z_{cp}^{sys}$ conditioned on the target speaker’s timbre $q_t^{trg}$. This clear separation—LLM for prosody and content, decoder for timbre—enables flexible zero-shot control of prosody and timbre.

\begin{table*}[htbp]
% \small
% \setlength{\tabcolsep}{0.5mm}
% \renewcommand\arraystretch{0.9}
\centering
\caption{Comparisons of various codec models for speech reconstruction on the LibriSpeech test-clean dataset. The WER is evaluated via a HuBERT-based ASR system. Bold values indicate the best for each token rate.}
\label{tab:codec_com}
\small
\setlength{\tabcolsep}{1mm}
\renewcommand\arraystretch{0.9}
% \resizebox{\textwidth}{!}{%
\begin{tabular}{lcccccccccc}
\toprule
Model & \makecell{Token\\Rate} & \makecell{Codebook\\Size} & \makecell{Disentanglement\\Ability} & WER $\downarrow$ & STOI $\uparrow$ & \makecell{PESQ\\WB$\uparrow$} & \makecell{PESQ\\NB$\uparrow$} & \makecell{SSIM$\uparrow$} & \makecell{UT\\MOS$\uparrow$} \\
\midrule
Ground Truth & - & - & -  & 1.96 & 1.00 & 4.64 & 4.55 & 1.00 & 4.09 \\
\midrule
BigCodec & 80 & 8192 & - & \textbf{2.76} & \textbf{0.93} & \textbf{2.68} & \textbf{3.27} & \textbf{0.84} & \textbf{4.11} \\
%\midrule
WavTokenizer & 75 & 4096 & - & 3.98 & 0.90 & 2.13 & 2.63 & 0.65 & 3.79 \\
Encodec & 75 & 1024 & - & 28.92 & 0.77 & 1.23 & 1.48 & 0.25 & 1.25 \\
MSR-Codec-424 &62.5 &500/32/64 & $\checkmark$ &- &0.84 &2.37 &1.82 &0.80 &4.15  \\
\midrule
DAC & 50 & 1024 & - & 74.55 & 0.62 & 1.06 & 1.20 & 0.08 & 1.25 \\
SpeechTokenizer & 50 & 1024 &- & 5.01 & 0.64 & 1.14 & 1.30 & 0.17 & 1.27 \\
Mimi & 50 & 2048 & - & 4.89 & 0.85 & 1.64 & 2.09 & 0.50 & 3.03 \\
StableCodec & 50 & 15625 & - & 5.12 & 0.91 & 2.24 & 2.91 & 0.62 & \textbf{4.23} \\
X-codec & 50 & 1024 & - & 3.42 & 0.83 & 1.84 & 2.38 & 0.52 & 4.05 \\
X-codec2 & 50 & 65536 & - & \textbf{2.47} & \textbf{0.92} & 2.43 & 3.04 & \textbf{0.82} & 4.13 \\
BiCodec & 50 & 8192 & - & -  & \textbf{0.92} & 2.51  & \textbf{3.13}   & 0.80   & 4.18 \\
\midrule
DisCodec & 50 & 65536 & $\checkmark$ & 2.92 & 0.86 & 1.98  & 2.33   & 0.81   & 4.10  \\
\bottomrule
\vspace{-15pt}
\end{tabular}
% }
\end{table*}

\section{Experiment}
\vspace{-5pt}

\subsection{Experimental Settings}
\label{tab:seting}

\textbf{Training Set.}
For DisCodec training, we utilize a 26k-hour mixed corpus (16-24kHz) curated from internal sources to ensure diversity in speakers and speaking prosody. In Stage 1, all speech samples are resampled to 16kHz for decoupling space learning, while only 24kHz samples are utilized in Stage 2 to ensure high-quality reconstruction.
For both stages, 80-dimensional Mel-spectrograms are extracted with a 50 ms frame length and 20 ms frame shift for the prosody and timbre branches.
Regarding the Text-to-Codec LM, a 120k-hour speech corpus, comprising Emilia~\cite{emilia} and the aforementioned internal dataset, is employed for pretraining. 
Subsequently, the LM undergoes SFT on a selected 5k-hour 24kHz subset.

To further enhance the performance of both the DisCodec and the Text-to-Codec LM, we conducted a fine-tuning stage using an additional 6k-hour high-quality 24kHz speech dataset.

\textbf{Configuration.}
DisCodec is trained for 500k steps with a total batch size of 176 on 8 NVIDIA A800 GPUs. We use the Adam optimizer ($lr=1e^{-4}$, $\beta_1=0.5$, $\beta_2=0.9$) with a linear warm-up for the first 5k steps. The codebook sizes for the FSQ layers are set to 65,536 (content), 46,656 (prosody/timbre), and 65,536 (content-prosody). The sequence length of the global timbre token is 48.
For the Text-to-Codec LM, initialized from the Qwen2.5-1.5B model\footnote{\url{https://huggingface.co/Qwen/Qwen2.5-1.5B}}, training runs for 8 epochs on 8×A800 GPUs, utilizing AdamW ($lr=2e^{-4}$, $\beta_1=0.9$, $\beta_2=0.98$).

\textbf{Evaluation Settings.}
For codec, following previous studies~\cite{sparktts}, the LibriSpeech \textit{test-clean} subset~\cite{librispeech-test-clean} is used for reconstruction assessment. Additionally, 1,680 conversion pairs (60 expressive source speeches \& 28 target speakers from SEED-TTS-Eval~\cite{seedtts}) are employed to verify the decoupling capability of the codec. 
For controllable generation, the widely used SEED-TTS-Eval (test-zh \& test-en) is adopted for evaluating voice cloning. 
We utilize a self-built prosody set for prosody control evaluation. 
We compare our method comprehensively against state-of-the-art baselines in three key areas: speech codec quality, disentanglement capability via voice conversion, and zero-shot TTS performance. 
For details of all compared models, please refer to Appendix~\ref{ap} and~\ref{ap2}.

\textbf{Evaluation Metrics.}
\textit{Objective Metrics}: We evaluate codec quality using PESQ~\cite{pesq} and UTMOS~\cite{utmos}. 
Speech intelligibility is measured via STOI~\cite{stoi}, Word Error Rate (WER), and Character Error Rate (CER), where WER and CER are computed using Whisper-large-v3~\cite{whisper} and Paraformer~\cite{paraformer}, respectively.
Speaker Similarity (SSIM) is calculated by a Speaker Verification (SV) model~\cite{wavellmlarge} to assess timbre consistency.
F0 Correlation Coefficient (F0$_{\text{cor}}$) is used to assess F0 contour consistency. 
\textit{Subjective Metrics}: An AB preference test~\cite{diclet} is adopted to subjectively compare samples synthesized by two models, where participants are asked to select the sample that sounds closer to the reference in terms of speaker timbre or prosody.

\vspace{-5pt}
\subsection{Experimental Results}
\subsubsection{Performance of DisCodec}

\textbf{Reconstruction Performance.} 
Table~\ref{tab:codec_com} presents the comparison of codec reconstruction performance among various neural codecs on the LibriSpeech \textit{test-clean} dataset~\cite{librispeech-test-clean}. In terms of perceptual quality, DisCodec achieves a UTMOS of 4.10, positioning it within the top tier of models at this token rate. 
DisCodec achieves an SSIM score of 0.81 in speaker similarity, which is on par with the top-performing X-Codec2.
Additionally, DisCodec demonstrates comparable results in content preservation with a WER of 2.92\%.
These results demonstrate that our proposed DisCodec achieves highly competitive and well-balanced performance at 50 tokens/s, despite its primary design focus on disentanglement.

\begin{table}[ht]
\centering
\setlength{\tabcolsep}{2.0pt}
\caption{Objective evaluation of zero-shot VC.}
\label{tab:obj_vc}
\small
\setlength{\tabcolsep}{3mm}
\renewcommand\arraystretch{1.0}
\begin{tabular}{lccc}
\toprule
Model & UTMOS$\uparrow$ & SSIM$\uparrow$ & F0$_{cor}$ $\uparrow$ \\
\midrule
CosyVoice2 &3.95 &0.55 &0.48  \\
Vevo       &4.0 &0.60 &0.50 \\
SeedVC     &\textbf{4.04} &0.58 &0.56 \\
\midrule
\textit{DisCodec}    &3.98 &\textbf{0.61} &\textbf{0.59} \\
% tyzr 0056
\bottomrule
\vspace{-15pt}
\end{tabular}
\end{table}

\textbf{Disentanglement Evaluation.}  In DisCodec, speech is disentangled into content-prosody and speaker timbre representations. To evaluate the effectiveness of this disentanglement, we conduct evaluations on the zero-shot voice conversion (VC) task, which requires models to convert speaker timbre while preserving the prosody and content of the source speech. As shown in Table~\ref{tab:obj_vc}, DisCodec achieves the highest target timbre similarity (SSIM) and F0 correlation (F0$_{\text{cor}}$), demonstrating its superior capability in simultaneous timbre conversion and prosody preservation. Moreover, this effective disentanglement does not compromise naturalness, with UTMOS scores competitive with the best baselines. The VC results confirm that DisCodec's representations achieve a superior balance between disentanglement and reconstruction.

\begin{figure}[t]
    \centering    
    \vspace{0.5em} 
    
    % 第二行 - 2个子图
    \begin{subfigure}[b]{0.49\linewidth}
        \centering
        \includegraphics[width=\linewidth]{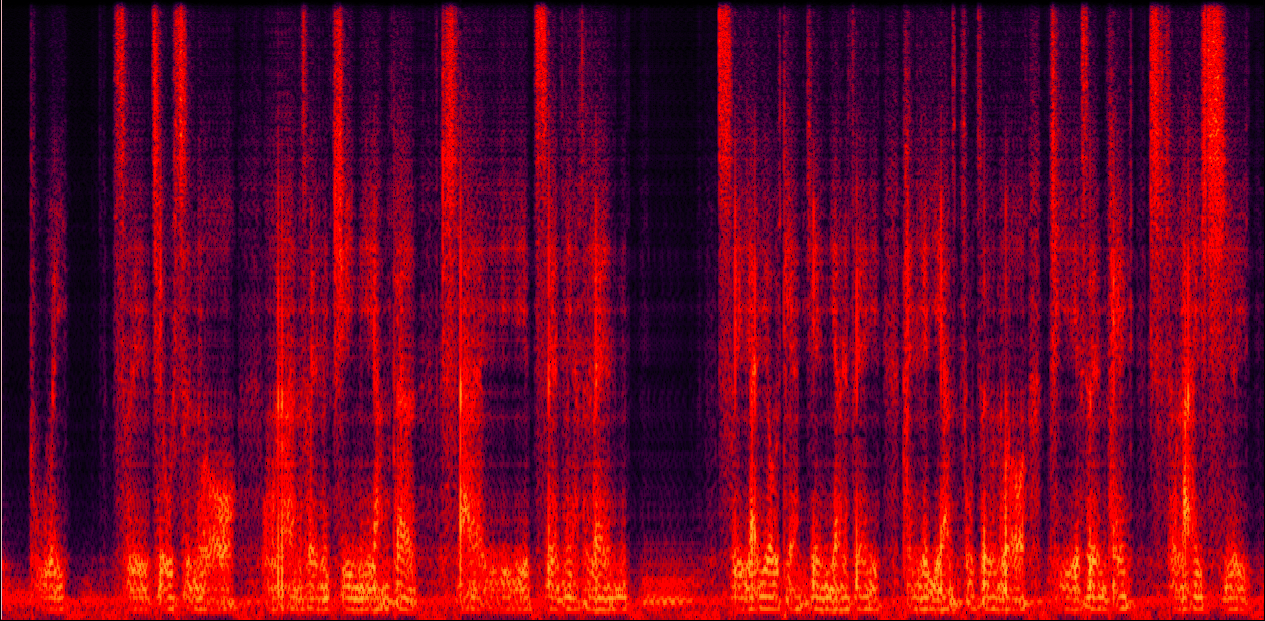}
        \caption{Content-only}
        \label{fig:sub_content}
    \end{subfigure}
    \hfill
    \begin{subfigure}[b]{0.49\linewidth}
        \centering
        \includegraphics[width=\linewidth]{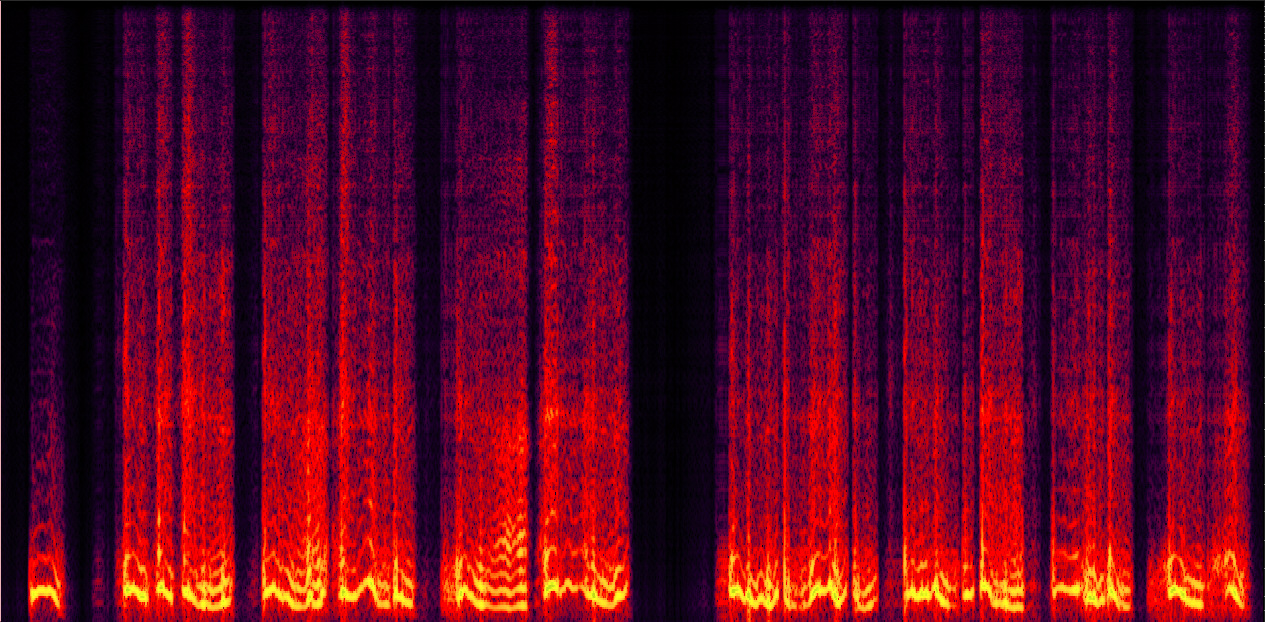}
        \caption{Prosody-only}
        \label{fig:sub_prosody}
    \end{subfigure}
    
     \vspace{0.5em} % 行间距，可以调整
    
    % 第三行 - 2个子图
    \begin{subfigure}[b]{0.49\linewidth}
        \centering
        \includegraphics[width=\linewidth]{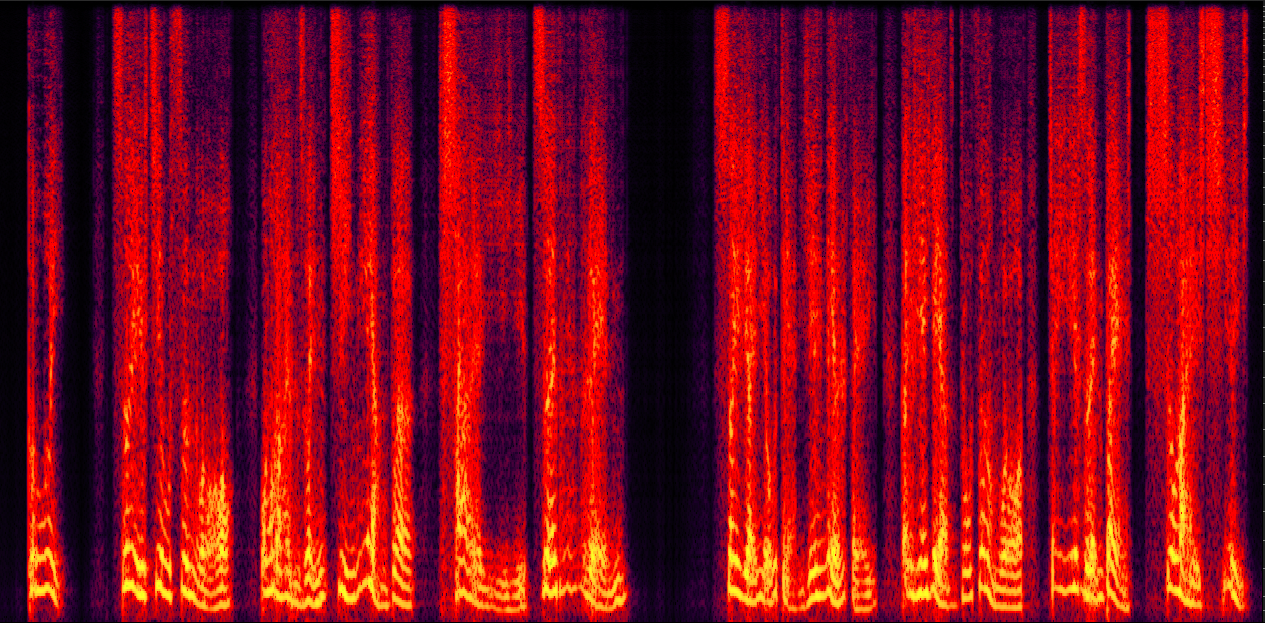}
        \caption{Content+Prosody}
        \label{fig:sub_cp}
    \end{subfigure}
    \hfill
    \begin{subfigure}[b]{0.49\linewidth}
        \centering
        \includegraphics[width=\linewidth]{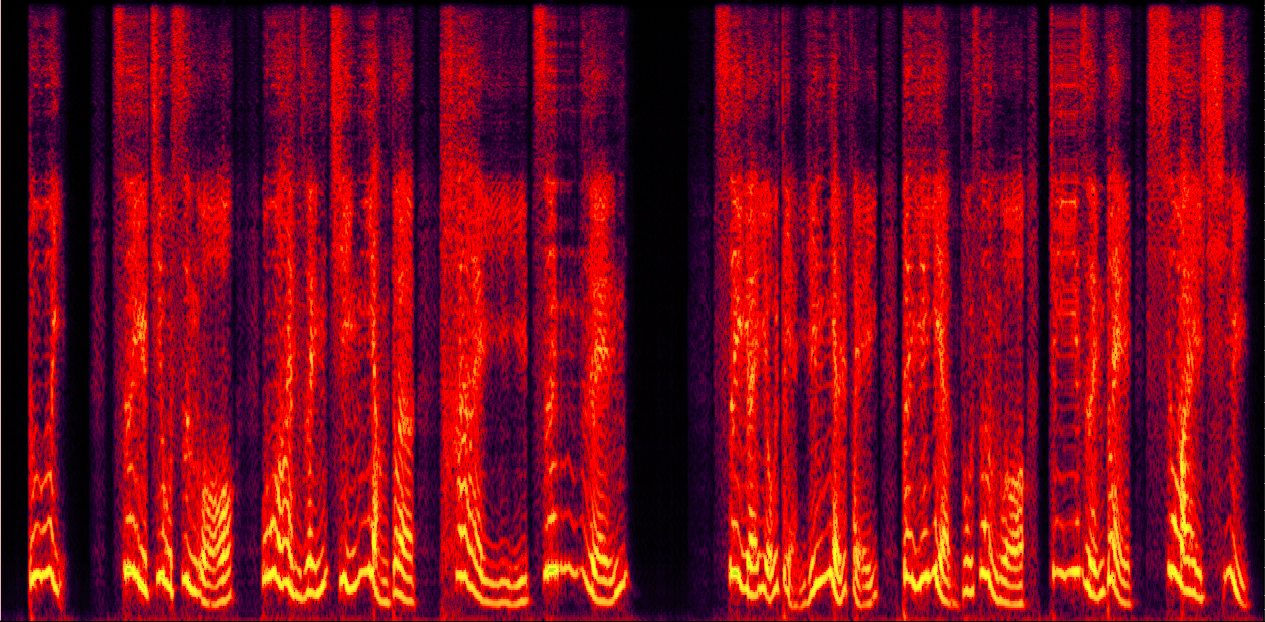}
        \caption{Full Reconstruction}
        \label{fig:sub_full}
    \end{subfigure}

    \caption{Disentanglement visualization.}
    \label{fig:dispng}
    \vspace{-15pt}
\end{figure}

\textbf{Visual Analysis.} We further visualize the disentanglement capability by performing reconstruction using different combinations of decoupled representations, forming four modes: content-only~(\textit{$q_c$}), prosody-only~(\textit{$q_p$}), content+prosody~(\textit{$q_c+q_p$}), and full reconstruction~(\textit{$q_c+q_p+q_t$}). The results are shown in Fig.~\ref{fig:dispng}. 
Using content $q_{c}$ alone (Fig.~\ref{fig:dispng}a) preserves phonemes but lacks F0 and harmonic details, while Fig.~\ref{fig:dispng}b exhibits the intensity and rhythm of prosody without linguistic intelligibility. 
When integrating content and prosody~(Fig.~\ref{fig:dispng}c), the formant structure aligns closely with the linguistic content. 
Finally, timbre information further adjusts the speaking expression to match the target speaker~(Fig.~\ref{fig:dispng}d). 
This visualization confirms the robust disentanglement performance of DisCodec. 

Due to space constraints, ablation studies on hierarchical prosody modeling and the soft orthogonality constraint are detailed in Appendix~\ref{ablation}.

\subsubsection{Controllability of DisCo-Speech}
In this section, we evaluate the system's capabilities in zero-shot prosody control and voice cloning.

\textbf{Zero-shot Timbre and Prosody Control.} To quantitatively evaluate independent controllability, we conduct an AB preference test. 
The test involves 10 expert listeners.
We construct an evaluation set of 1,000 prosody-timbre prompt pairs (deliberately sourced from different speakers), each combined with 5 distinct texts, yielding 5,000 generated utterances from which 500 samples are randomly selected for comparison.
As presented in Table~\ref{tab:contro_tts}, DisCo-Speech demonstrates superior performance in \textit{Style} scenario, achieving higher preference rates in both timbre and prosody. 
This confirms that for stylistic attributes (e.g., storytelling, poetry), which are primarily manifested through rhythm and speech rate, DisCo-Speech effectively captures the target prosody while robustly maintaining the speaker's timbre.
As observed from the generated samples\footnote{\url{https://disco-speech.github.io/DisCo-demo/}}, DisCo-Speech also demonstrates stability in \textit{cross-gender prosody transfer}.

In the \textit{Emotion} scenario, we observe a trade-off that highlights architectural differences. 
Index-TTS 2 slightly outperforms DisCo-Speech in prosody preference due to its high expressiveness; however, it lags in timbre consistency. 
We observe that Index-TTS 2's strong emotional expressiveness often comes at the cost of source timbre leakage. Since intense emotions entail intrinsic timbre variations~\cite{taslpemo}, Index-TTS 2 tends to entangle the speaker's timbre with emotional expressiveness.
In contrast, DisCo-Speech enforces stricter disentanglement: the LM captures prosody from the prompt, and the DisCodec decoder renders the target timbre. This ensures that the target speaker's timbre remains uncompromised even during intense emotion transfer.

\begin{table}[ht]
\centering
\caption{AB preference test results. We evaluate speaker timbre consistency and prosody similarity on both \textit{Emotion} and \textit{Style} scenarios.}
\label{tab:contro_tts}
\small
\setlength{\tabcolsep}{0.8mm}
\renewcommand\arraystretch{1.25}
\begin{tabular}{l c c c c}
\toprule
\multirow{2}{*}{\textbf{Scenario}} & \multirow{2}{*}{\textbf{Aspect}} & \multicolumn{3}{c}{\textbf{Preference (\%)}} \\
\cmidrule(lr){3-5}
& & \textbf{DisCo-Speech} & No Preference & \textbf{Baseline} \\
\midrule
\multicolumn{5}{c}{\textit{DisCo-Speech vs. Vevo}} \\
\midrule
\multirow{2}{*}{Emotion} & Timbre  & \textbf{45.3} & 14.5 & 40.2 \\
                         & Prosody & \textbf{50.6} & 12.7 & 36.7 \\
\cmidrule{1-5}
\multirow{2}{*}{Style}   & Timbre  & \textbf{51.5} & 27.2 & 21.3 \\
                         & Prosody & \textbf{48.9} & 31.1 & 20.0 \\
\midrule
\multicolumn{5}{c}{\textit{DisCo-Speech vs. IndexTTS 2}} \\
\midrule
\multirow{2}{*}{Emotion} & Timbre  & \textbf{42.5} & 23.8 & 33.7 \\
                         & Prosody & 36.0 & 25.5 & \textbf{38.5} \\
\cmidrule{1-5}
\multirow{2}{*}{Style}   & Timbre  & \textbf{37.4} & 33.6 & 29.0 \\
                         & Prosody & \textbf{41.7} & 25.3 & 33.0 \\
\bottomrule
\end{tabular}
\end{table}

\textbf{Voice Cloning.} 
Following the evaluation protocol of the SEED-TTS-Eval~\cite{seedtts} benchmark, we assess the voice cloning capability of DisCo-Speech against state-of-the-art TTS models. As shown in Table~\ref{tab:clone_ob}, DisCo-Speech maintains high speech intelligibility in both English and Mandarin.
In terms of speaker similarity, it achieves performance on par with other one-stage autoregressive models such as Spark-TTS, which also employ an LM and codec decoder architecture, while performing slightly below flow-matching-based systems that utilize powerful yet complex pipelines.
In summary, results from both prosody control and voice cloning validate DisCo-Speech as an effective and integrated framework for high-quality zero-shot controllable speech generation.

\begin{table}[htbp]
\centering
\setlength{\tabcolsep}{12pt}
\small
\caption{Results of voice cloning. $\clubsuit$ marks the systems which 
supports independent control of timbre and prosody.}
\label{tab:clone_ob}
\small
% \small
\setlength{\tabcolsep}{0.5mm}
\renewcommand\arraystretch{0.95}
\newcommand{\club}{\textsuperscript{$\clubsuit$}}
\begin{tabular}{lccccc}
\toprule
\multirow{2}{*}{Model} & \multirow{2}{*}{Params} & \multicolumn{2}{c}{Test-EN} & \multicolumn{2}{c}{Test-ZH}  \\
\cmidrule(lr){3-4} \cmidrule(lr){5-6}
 & & WER $\downarrow$ & SSIM $\uparrow$ & CER $\downarrow$ & SSIM $\uparrow$ \\
\midrule
\multicolumn{6}{c}{\textit{Multi-Stage or NAR Methods}} \\
\midrule
F5-TTS & 0.3B &1.83 &0.647 &1.56 & 0.741 \\
CosyVoice2  & 0.5B & 2.57 & 0.652 & 1.45 & 0.748\\
%CosyVoice3  & 1.5B & 2.21 & 78.9 & 1.12 & 83.7\\
Index-TTS 2\club & 1.5B & 2.23 & 0.706 & 1.03 & 0.765\\
FireRedTTS & - & 3.82 & 0.460 & 1.51 & 0.635 \\
Vevo\club   & - &2.53 &0.664 &3.99 &0.723 \\
\midrule
\multicolumn{6}{c}{\textit{One-Stage AR Methods}} \\
\midrule
Llasa-1B    &1B &3.22 &0.572 &1.89 &0.669 \\
Spark-TTS   &0.5B & 1.98 & 0.584 & 1.20 & 0.672  \\
\textit{DisCo-Speech}\club  & 1.5B &3.01  &0.597 &2.08 & 0.677 \\
\bottomrule
\end{tabular}
\vspace{-10pt}
\end{table}

\section{Conclusions}
\vspace{-5pt}
We presented DisCo-Speech, a novel framework for zero-shot controllable speech generation that achieves independent control over speaker timbre and speaking prosody. At its core lies DisCodec, a disentangled speech codec that explicitly factorizes speech into content, prosody, and timbre subspaces through a principled two-stage training paradigm. This design effectively resolves the inherent trade-off between disentanglement and reconstruction while producing downstream-friendly representations for standard LM. With this design, given a prosody template, the LM executes prosodic continuation based on the text, and the DisCodec decoder further injects the desired speaker timbre. Extensive experiments demonstrate the superior controllability of DisCo-Speech in both voice cloning and prosody control.

\newpage
\section{Limitations}  
Despite its promising performance, DisCo-Speech has certain limitations. First, speaker similarity remains relatively lower than that of multi-stage systems, likely due to the inherent variability of autoregressive generation~\cite{sparktts} and the compact nature of codec representations~\cite{recent}. 
Additionally, while trained on extensive data, the quality and expressive diversity of the training corpus remain constrained, which may lead to instability when generating highly exaggerated speaking prosody. 
Furthermore, maintaining the delicate balance between disentanglement purity and reconstruction fidelity presents an ongoing challenge, as enhancing disentangling may sometimes come at the cost of fine-grained acoustic details.
In future work, we aim to improve performance through higher-quality datasets and advanced architectural designs that jointly optimize both disentanglement effectiveness and detailed reconstruction capability.

\bibliography{custom}

\appendix

\section{Training Objectives and Hyperparameters of DisCodec}
\label{aploss}

\subsection{Training Objectives}

The training of DisCodec is divided into two stages. 
Below we detail the total loss functions and their specific purposes for each stage.

\paragraph{Stage 1: Tri-factor Disentanglement.}
In stage 1, the goal is to disentangle speech into content, prosody, and timbre while ensuring the integrity of each attribute. 
The objective $\mathcal{L}_{Stage1}$ is a weighted sum of reconstruction losses and disentanglement constraints:

\begin{equation}
\label{eq:total_s1}
\begin{aligned}
    \mathcal{L}_{Stage1} = & \lambda_{rec}\mathcal{L}_{rec} + \lambda_{pho}\mathcal{L}_{pho} + \lambda_{spk}\mathcal{L}_{spk} \\
    & + \lambda_{f0}\mathcal{L}_{f0} + \lambda_{cor}\mathcal{L}_{cor} + \lambda_{grl}\mathcal{L}_{grl.spk} \\
    & + \lambda_{soft}(\mathcal{L}_{soft}^{p,c} + \mathcal{L}_{soft}^{p,t}),
\end{aligned}
\end{equation}

where:
\begin{itemize}
    \item $\mathcal{L}_{rec}$: Combines the L1 loss in the time domain and the multi-scale Mel-spectrogram loss~\cite{dac}  ($\mathcal{L}_{mel}$) to ensure waveform reconstruction.
    \item $\mathcal{L}_{pho}$: The Cross-Entropy loss calculated between the predicted phone probabilities from the content representation $q_c$ and the extracted labels from the Wav2Vec-based~\cite{wav2vec} recognizer.
    \item $\mathcal{L}_{spk}$: The Cross-Entropy loss for speaker identification based on global timbre representation $q_t$.
    \item $\mathcal{L}_{f0}$: The L2 regression loss between the F0 predicted by the first prosody FSQ layer and the ground truth F0.
    \item $\mathcal{L}_{grl.spk}$: The Cross-Entropy loss for the GRL-based~\cite{grl} speaker classifier applied to the prosody branch, aiming to prevent timbre leakage in $q_p$.
    \item $\mathcal{L}_{cor}$ and $\mathcal{L}_{soft}$: The correlation and soft orthogonality constraints as defined in Eq.~\ref{eq2} and Eq.~\ref{eq34}, governing the disentanglement optimization.
\end{itemize}

\paragraph{Stage 2: Fusion and Reconstruction.}
In stage 2, the decoder is updated with a GAN-based objective to improve perceptual quality, following the BigVGANv2 setup. 
The stage 2 loss $\mathcal{L}_{stage2}$ is defined as:

\begin{equation}
    \mathcal{L}_{stage2} = \lambda_{mel}\mathcal{L}_{mel} + \lambda_{fm}\mathcal{L}_{fm} + \lambda_{adv}\mathcal{L}_{adv},
\end{equation}
where $\mathcal{L}_{mel}$ is the Mel-spectrogram reconstruction loss, $\mathcal{L}_{fm}$ is the feature matching loss computed from the intermediate layers of the discriminator, and $\mathcal{L}_{adv}$ is the adversarial loss derived from the multi-scale discriminators~\cite{bigvganv2}.
Specifically, we apply LayerNorm to the quantized embeddings ($q_{p1}$ and $q_{p2}$ in Eq.~\ref{eq2}) before computing cosine similarities to stabilize training.

\subsection{Hyperparameter Settings}

The detailed hyperparameter configurations for the loss weights ($\lambda$) and the specific coefficients used in DisCodec disentanglement constraints ($\alpha, \beta$) are listed in Table~\ref{tab:hyperparams}.

\begin{table}[h]
\centering
\small
\caption{Detailed hyperparameters and loss weights used in DisCodec training.}
\label{tab:hyperparams}
\renewcommand\arraystretch{1.2}
\begin{tabular}{l c c}
\toprule
\textbf{Hyperparameter} & \textbf{Notation} & \textbf{Value} \\
\midrule
\multicolumn{3}{c}{\textit{Stage 1: Tri-factor Disentanglement}} \\
\midrule
Reconstruction Weight & $\lambda_{rec}$ & 12.5 \\
Phonetic Loss Weight & $\lambda_{pho}$ & 2.0 \\
Speaker Loss Weight & $\lambda_{spk}$ & 1.0 \\
F0 Loss Weight & $\lambda_{f0}$ & 1.5 \\
GRL-speaker Loss Weight & $\lambda_{grl.spk}$ & 0.1 \\
Correlation Loss Weight & $\lambda_{cor}$ & 0.5 \\
Soft Constraint Weight & $\lambda_{soft}$ & 5.0 \\
Correlation Target & $\alpha$ & 0.2 \\
P-C Decoupling Coeff. & $\beta_c$ & 0.01 \\
P-T Decoupling Coeff. & $\beta_t$ & $1\times10^{-4}$ \\
\midrule
\multicolumn{3}{c}{\textit{Stage 2: Fusion \& Reconstruction}} \\
\midrule
Mel-Reconstruction Weight & $\lambda_{mel}$ & 15.0 \\
Feature Matching Weight & $\lambda_{fm}$ & 1.0 \\
Adversarial Weight & $\lambda_{adv}$ & 1.0 \\
\bottomrule
\end{tabular}
\end{table}

On the choice of $\beta_c$ and $\beta_t$ in Soft Orthogonality Constraints: The hyperparameters $\beta_c$ and $\beta_t$ in Eqs. (3) and (4) represent the target cosine similarity between the prosody subspace and the content/timbre subspaces.
\textbf{Content-Prosody Relaxation} ($\beta_c = 0.01$): Unlike timbre, which is globally static, prosody and content are both temporal-dynamic and intrinsically coupled (as shown in Fig.~\ref{fig:dispng}) in human speech (e.g., lexical stress and tonal alignment in Mandarin~\cite{diclet}). 
Imposing a stringent constraint would force the model to strip essential fine-grained prosodic information, resulting in loss of prosodic detail, which inevitably leads to degradation in both prosodic expressiveness and overall naturalness.
We set $\beta_c$ to a small but non-zero value (0.01) to allow a "soft" overlap, ensuring that the content tokens retain sufficient alignment information for the LM to predict accurate prosody.
\textbf{Timbre-Prosody Strictness} ($\beta_t = 1 \times 10^{-4}$): Speaker timbre and prosody are theoretically more independent—a target speaker's timbre characteristics should remain consistent regardless of emotional states or speaking styles~\cite{taslpemo,copycat}. 
To prevent "timbre leakage" into the prosody embedding (which would cause the timbre to change when the style changes), we enforce a near-orthogonal constraint. 
This ensures that $q_p$ captures a timbre-agnostic prosody representation.
Based on empirical observations from experiments on our dataset, we found that significantly deviating from this ratio led to a decline in both reconstruction quality and the stability of independent attribute control.

\begin{table*}[htbp]
\centering
\small
\caption{Ablation study on the DisCodec core components evaluated on Zero-shot VC.}
\label{tab:ablation_all}
\begin{tabular}{l c c c}
\toprule
\textbf{Model Variant} & \textbf{UTMOS} $\uparrow$ & \textbf{F0 Corr} $\uparrow$ & \textbf{SSIM} $\uparrow$ \\
\midrule
\multicolumn{4}{c}{\textit{ Hierarchical Prosody Modeling}} \\
\midrule
\quad w/o Residual & 3.90 & 0.62 & 0.59 \\
\quad w/o $\mathcal{L}_{cor}$ & 3.81 & 0.48 & 0.53 \\
\midrule
\multicolumn{4}{c}{\textit{ Disentanglement Constraint}} \\
\midrule
\quad No Constraint ($\lambda_{soft}=0$) & 3.91 & \textbf{0.64} & 0.48 \\
\quad Hard Constraint ($\beta_{c}, \beta_{t} = 0$) & 3.83 & 0.52 & \textbf{0.69} \\
\midrule
Proposed & \textbf{3.98} & 0.59 & 0.61 \\
\bottomrule
\end{tabular}
\end{table*}

\section{Additional Ablation Studies of DisCodec}
\label{ablation}

In this section, we provide ablation studies on the zero-shot voice conversion (VC) task to validate the two core designs in DisCodec: 
\begin{itemize} 
\item \textbf{Hierarchical Prosody Modeling:} The prosody encoder in DisCodec employs a dual-layer residual FSQ structure supervised by a correlation loss $\mathcal{L}_{cor}$. 
The first layer explicitly models F0, while the second residual fsq layer captures prosodic related information beyond pitch. 
To verify the necessity of this design, we compare the proposed method with two variants: 
1) \textbf{w/o Residual}: Only using the first FSQ layer (supervised by F0 loss) to represent prosody; 
2) \textbf{w/o $\mathcal{L}_{cor}$}: Using the dual-layer structure but removing the correlation constraint.
\item \textbf{Disentanglement Constraint:} Achieving disentanglement involves a trade-off between attribute purity and information preservation. We compare our proposed \textbf{Soft Constraint} strategy against two extremes: 
1) \textbf{No Constraint} ($\lambda_{soft}=0$), where no penalty is applied to the dependencies between attributes; 
2) \textbf{Hard Constraint}, which enforces strict orthogonality ($\beta_{c}, \beta_{t} = 0$) between attribute representations.
\end{itemize}
Evaluation is conducted on the zero-shot VC task, focusing on timbre similarity (SSIM), speech quality (UTMOS), and F0 Correlation (F0 Corr).

As shown in Table~\ref{tab:ablation_all}, for prosody modeling, the \textbf{w/o Residual} variant yields a slightly higher F0 Corr, confirming its specialization in pitch tracking. 
However, its drops in UTMOS suggest that a pitch-only representation ignores broader prosodic nuances, which are essential for naturalness.
Removing the correlation loss (w/o $\mathcal{L}_{cor}$) results in the lowest UTMOS among all variants, indicating that without proper guiding the residual layer, it fails to learn complementary prosodic information and instead introduces harmful noise that degrades speech reconstruction.

Regarding disentanglement, the \textbf{No Constraint} variant exhibits the lowest SSIM, indicating severe timbre leakage into prosodic representation. 
Conversely, the \textbf{Hard Constraint} achieves the highest SSIM but suffers from a degraded UTMOS. 
This indicates that excessive disentanglement leads to a loss of detailed acoustic information. 
Our \textbf{Soft Constraint} strikes the optimal balance, maintaining high timbre consistence while preserving naturalness. 

\section{Compared Codec Methods}
\label{ap}

\begin{itemize}

\item BigCodec~\cite{bigcodec}: A VQ-based single-stream codec for speech.

\item Encodec~\cite{encodec}: An RVQ-based codec designed for universal audio compression.

\item DAC~\cite{dac}: An RVQ-based codec for universal audio.

\item Mimi~\cite{mimi}: An RVQ-based codec with semantic constraint for speech.

\item Single-Codec~\cite{singlecodec}: A single-stream Mel codec that incorporates speaker embeddings. The reconstruction results for this method are provided by the authors.

\item SpeechTokenizer~\cite{speechtokenizer}: An RVQ-based codec with semantic distillation for speech.  

\item X-codec~\cite{xcodec1}: An RVQ-based codec with semantic distillation for speech.  

\item X-codec2~\cite{xcodec}: A FSQ-based single-stream codec with semantic distillation for speech.  

\item StableCodec~\cite{stablecodec}: A residual FSQ-based tokenizer for speech.  

\item MSR-Codec~\cite{msrcodec}: A residual VQ-based tokenizer for speech disentanglement.  

\item WavTokenizer~\cite{wavtokenizer}: A single VQ codebook-based tokenizer for universal audio.

\item BiCodec~\cite{sparktts}: A single-stream speech codec that decomposes speech into two complementary token types: low-bitrate semantic tokens for linguistic content and fixed-length global tokens for speaker attributes.

\end{itemize}

\section{Compared Zero-shot Methods}
\label{ap2}

\begin{itemize}

\item CosyVoice2~\cite{cosyvoice2}: A two-stage model with an LM for semantic tokens and flow matching for acoustic features generation.

\item Vevo~\cite{vevo}: A two-stage model with SSL-based disentangled speech codec for prosody and timbre control generation.

\item SeedVC~\cite{seedvc}: A flow matching-based VC methods with timbre perturbation.

\item Spark-TTS~\cite{sparktts}:  A single-stage model that uses a single-codebook speech codec coupled with an LLM and codec decoder for speech generation.

\item Llasa~\cite{llasa}: A single-stream codec-based TTS model that uses a single AR language model for code prediction.

\item FireRedTTS~\cite{fireredtts}: A two-stage model similar to Seed-TTS, using an AR LM for semantic tokens and flow matching for acoustic features.

\item Index-TTS2~\cite{indextts2}: A two-stage model with emotion disentanglement and duration control.

\item F5-TTS~\cite{f5tts}: A flow matching-based method that also uses Mel spectrograms as acoustic features.

\end{itemize}

\end{document}